\documentclass[aps,prd,twocolumn,showpacs,amssymb,floatfix]{revtex4}
\newcommand{\be}{\begin{equation}}
\newcommand{\ee}{\end{equation}}
\newcommand{\beq}{\begin{eqnarray}}
\newcommand{\eeq}{\end{eqnarray}}
\usepackage{bm}
\usepackage{graphicx}%
\usepackage{epsfig}
\usepackage{amsmath}
\usepackage{amsfonts,amssymb}
\usepackage{hhline}
\usepackage{multirow}

\newcommand{\dslash}{\not{\hbox{\kern-2pt $\partial$}}}

\begin{document}
\title{Conformal vs confining scenario in SU(2) with adjoint fermions}
\author{L.~Del~Debbio$^a$, B.~Lucini$^b$, A.~Patella$^b$, C.~Pica$^a$ and A.~Rago$^b$}
\affiliation{$^a$ SUPA, School of Physics and Astronomy, University of Edinburgh, Edinburgh EH9 3JZ, UK}
\affiliation{$^b$ School of Physical Sciences, Swansea University, Singleton Park, Swansea SA2 8PP, UK}
\begin{abstract}
  The masses of the lowest--lying states in the meson and in the gluonic sector of an SU(2) gauge
  theory with two Dirac flavors in the adjoint representation are measured on the lattice at a
  fixed value of the lattice coupling $\beta = 4/g_0^2 = 2.25$ for values of the bare fermion mass
  $m_0$ that span a range between the quenched regime and the massless limit, and for various
  lattice volumes. Even for light constituent fermions the lightest glueballs are found to be
  lighter than the lightest mesons. Moreover, the string tension between two static fundamental
  sources strongly depends on the mass of the dynamical fermions and becomes of the order of the
  inverse squared lattice linear size before the chiral limit is reached. The implications of these
  findings for the phase of the theory in the massless limit are discussed and a strategy for
  discriminating between the (near--)conformal and the confining scenario is outlined.
\end{abstract}
\maketitle

The existence of a strongly--interacting sector beyond the Standard
Model that is responsible for the breaking of electroweak symmetry was
proposed many years
ago~\cite{Weinberg:1975gm,Susskind:1978ms}. This scenario is referred
to as Technicolor. The constituent fields of this sector describe
techniquarks and technigluons. The dynamically--generated mass scale
-- i.e.~the equivalent of the typical hadronic mass scale in QCD
-- separates the low--energy regime of the theory, of which the
Standard Model is an effective theory, from the regime in which the
technicolor sector becomes manifest; this scale is of the order of the
TeV. However, a simple rescaling of QCD to the TeV scale and
successive extensions to account for non--zero masses of Standard Model
fermions prove to be inadequate to describe the phenomenology of the
Standard Model itself (see {\it
e.g.}~\cite{Hill:2002ap,Sannino:2008ha} for a review). Walking
Technicolor provides a suggestive framework to cure some of these
problems~\cite{Holdom:1984sk,Yamawaki:1985zg,Appelquist:1986an}. In
this scenario, the technicolor theory behaves like QCD both in the
ultraviolet and in the infrared regimes; asymptotic freedom
characterizes the behavior at high--energies, while in the infrared
domain the theory is confining, and chiral symmetry is
broken. However, there is an intermediate range of energies in which
the dependence of the coupling on the energy scale is supposed to be
very mild. This is the walking regime, which should characterize
theories that are close to the conformal window. Although this
framework solves phenomenological problems of the original Technicolor
model, its most straightforward realization, as an SU($N$) gauge
theory with fundamental fermions, requires a large number of flavors to be compatible with electroweak
precision data~\cite{Peskin:1991sw}: the phenomenologically favored
theories with a low number of flavors $N_f$ and a low number of colors
$N$ have necessarily fermions in higher representations. 
Another interesting scenario, known as Conformal
Technicolor~\cite{Luty:2004ye}, assumes the existence of an underlying
theory with a strongly--interacting IR fixed point. Recent analytical
investigations using a wide range of techniques have tried to
characterize such (near--)conformal
theories~\cite{Ryttov:2007cx,Rattazzi:2008pe,Nunez:2008wi,Poppitz:2009uq,Armoni:2009jn}.

An interesting possibility for reconciling the Walking Technicolor
scenario with the experimental data is to consider gauge theories
coupled with fermions transforming in higher representations of the
gauge group~\cite{Sannino:2004qp}, in which the conformal phase (and
as a consequence the near--conformal phase) can be reached at values
of $N_f$ and $N$ that are not excluded by
phenomenology~\cite{Dietrich:2006cm}. In particular, the theory with
$N = N_f = 2$, and Dirac fermions in the adjoint representation is a likely
candidate for the realization of the (near--)conformal
scenario. Evidence for such (near--)conformal behavior is based on
analytical calculations performed relying on uncontrolled
approximations, or educated guesses. Hence, it is mandatory to
investigate the phase structure of those theories from first
principles. This has prompted several lattice calculations of bound
state
masses~\cite{Catterall:2007yx,DelDebbio:2008zf,Catterall:2008qk,Hietanen:2008mr,DeGrand:2008kx},
of the Dirac operator
spectrum~\cite{Fodor:2008hn,Fodor:2008hm,DeGrand:2009et} and of the
renormalization group flow~\cite{Shamir:2008pb,Hietanen:2009az,Bilgici:2009kh} of
candidate walking theories. These calculations have to be interpreted
with great care, since lattice systematic errors can obscure the
physical behavior. Nonetheless the picture emerging from these studies
is that the physics of gauge theories with fermions in the adjoint or
in the symmetric representation has a different signature than
QCD. Whether this is an indication of possible (near--)conformal
behavior or a manifestation of the limitations of the current
calculations is a question that can be answered only by gaining better
control on the chiral, the infinite volume, and the continuum limit
extrapolations. Interesting discussions of possible lattice signatures
have been presented in Refs.~\cite{DeGrand:2009mt,Luty:2008vs}.

Most of the spectrum--based studies have looked for signatures of
conformal or walking behavior in various observables; mesons have been
studied in
Refs.~\cite{Catterall:2007yx,DelDebbio:2008zf,Catterall:2008qk,Hietanen:2008mr,DeGrand:2008kx},
baryons in Ref.~\cite{Hietanen:2008mr} and Creutz ratios in
Ref.~\cite{Catterall:2008qk}. In a gauge theory with massless fermions
and chiral symmetry breaking, the vector meson has a mass of the order
of the dynamical scale of the theory $\Lambda$ and the pseudoscalar
meson is massless, since it is the Goldstone boson associated to the
spontaneous breaking of chiral symmetry. If the fermions have a small
mass $m \ll \Lambda$, the masses of the pseudoscalar and vector states
are given by
\begin{equation}
\label{eq:chisb1}
m_\mathrm{PS}  = a_\mathrm{PS} \sqrt{m} \ , \qquad 
m_\mathrm{V} = a_\mathrm{V} m + b_\mathrm{V} \ ,  
\end{equation}
where $b_\mathrm{V}$ is the mass of the vector meson in the chiral
limit. The other states in the spectrum are expected to behave like
the vector state. In the case of a conformal theory, at $m = 0$ bound
states cannot form. When a small mass is added, the theory is driven
away from conformality. Masses of bound states then scale as
\begin{equation}
\label{eq:conformal1}
m_\mathrm{PS}  = \alpha_\mathrm{PS} m^{\rho} \ , \qquad 
m_\mathrm{V} = \alpha_\mathrm{V} m^{\rho}  \ ,
\end{equation}  
where $\rho$ is related to the anomalous dimension of the mass $\gamma$ by
\begin{equation}
  \label{eq:critexp}
  \rho=1/(1+\gamma)\, .
\end{equation}


Since in ordinary Monte Carlo calculations it is impossible to
simulate directly the massless case, indications of
(near--)conformality can be sought using
Eqs.~(\ref{eq:chisb1}-\ref{eq:conformal1}). In particular, one would
measure the ratio $m_\mathrm{V}/m_\mathrm{PS}$ and check whether it
behaves like in QCD (i.e.~whether it goes to infinity as $m^{-1/2}$,
or it stays constant at least for a wide interval of masses before
diverging). The problem with this approach is that a constant ratio
$m_\mathrm{V}/m_\mathrm{PS} \simeq 1$ is also a characteristic of the
large fermion mass limit of SU($N$) gauge theories, and for a theory
that is not QCD (i.e.~for which we can not rely on guidance coming
from experiments) it is not clear a priori which bare mass would be
small enough for the onset of the chiral behavior to be
visible. Hence, there is a risk to confuse a confining, chiral
symmetry breaking theory with heavy fermions, and a (near--)conformal
theory in the infrared. The central point of this work is to show that
this ambiguity can be successfully resolved by comparing the spectrum
of the lightest mesons with gluonic observables.

In this note, we report on a numerical investigation of SU(2) gauge
theory with two flavors of Dirac fermions in the adjoint
representation for a fixed value of the lattice coupling $\beta$. With
this calculation, we aim to make a relevant step towards the
understanding of the chiral limit while keeping under control finite
size effects, but we will not be addressing the issue of the
extrapolation towards the continuum, our calculation being at fixed
lattice spacing. This point is crucial to understand the scope of the
conclusions that can be drawn from our calculations, since all our
results can potentially be affected by lattice artifacts, which could
distort the numerical results.  On the other hand, it is also worth
stressing that the results presented here explore a range of masses
much lighter than the ones in previous studies of this theory. At the
same time care is taken at taming finite volume effects at such small
masses.

In order to keep our presentation contained and accessible to a more
general audience, we defer the discussion of the lattice observables
and the presentation of our results in full to future
publications. Here, we limit our exposition to the essential aspects,
referring for the moment the interested reader to the quoted
literature for general discussions of the techniques. Simulations are
performed on a spacetime lattice with geometry $(2L) \times L^3$,
where $L=8,12,16$. The long direction plays the role of Euclidean
time, while the others are spatial directions.  Fermions have
antiperiodic boundary conditions in the temporal direction, and
periodic boundary conditions in the spatial directions. We use the
Wilson action for the gauge field and the Wilson discretization for
the Dirac operator (the details of the implementation are given in
Ref.~\cite{DelDebbio:2008zf}). The bare parameters are the bare
coupling $g_0$ and the bare mass in lattice units $a m_0$. The
coupling $g_0$ controls the size of the lattice spacing $a$. Our
calculation is performed at $\beta = 4/g_0^2 = 2.25$, which has been
found to be in the region of the phase diagram connected with the
continuum limit~\cite{Catterall:2008qk,Hietanen:2008mr}.

We measure masses of non--singlet mesons, the quark mass from the axial Ward identity
(PCAC mass), the masses of the $0^{++}$ and of the $2^{++}$ glueballs and the string
tension from the large distance exponential decay of correlators of
operators with the appropriate quantum numbers. More details on the
meson observables discussed in this work can be found in
Ref.~\cite{DelDebbio:2008zf}, which also contains a description of the
simulation techniques we use to generate the configurations, while for
measurements of quantities in the gluonic sector we follow
Ref.~\cite{Lucini:2004my}, from which we also borrow results for
gluonic observables in the SU(2) Yang--Mills theory. The data shown are
those obtained on the largest available lattice after checking that
results on the smaller lattices were compatible. Whenever this request
was not fulfilled we have discarded the data for the corresponding
observable. As a consequence, gluonic observables that are more
sensitive to finite--size effects could not be computed at the lighter
masses.

%
\begin{figure}[ht]
\includegraphics*[scale=0.35]{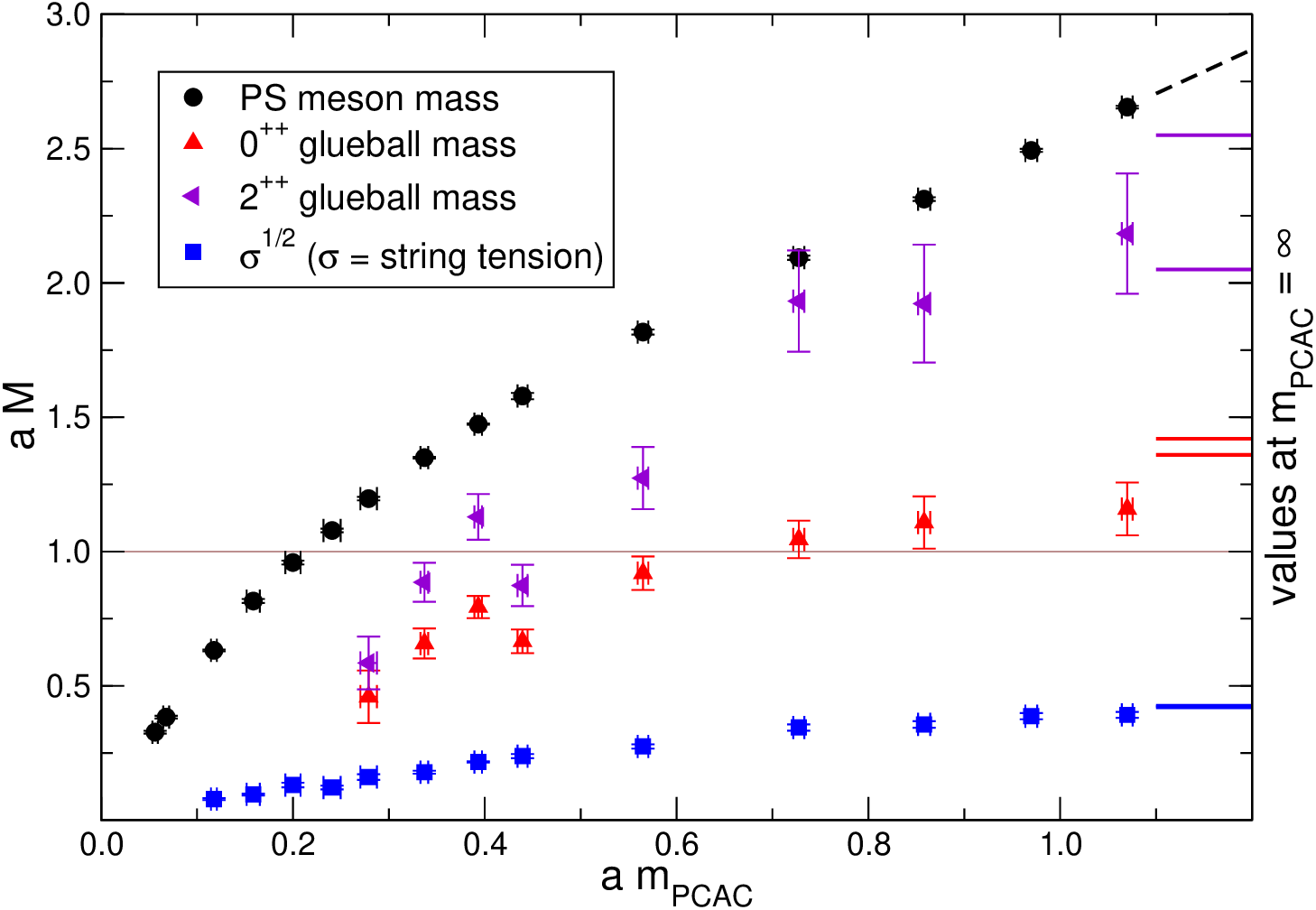}
\caption{Comparative plot of the various observables as a function of
  $m_\mathrm{PCAC}$. The lines at high PCAC mass show the quenched behavior
  of the various observables. The horizontal line at $a M = 1$ visually
  shows the separation between values of masses that are affected by
  lattice artifacts ($a M > 1$) and values for which the states are
  below the ultraviolet cutoff ($a M < 1$). Although ideally one
  wants all the states of interest to be free from lattice artifacts,
  due to the big separation of the scales, this condition is difficult
  to accommodate in practice.}
\label{fig:all}
\end{figure}

\begin{figure}[ht]
\includegraphics*[scale=0.34]{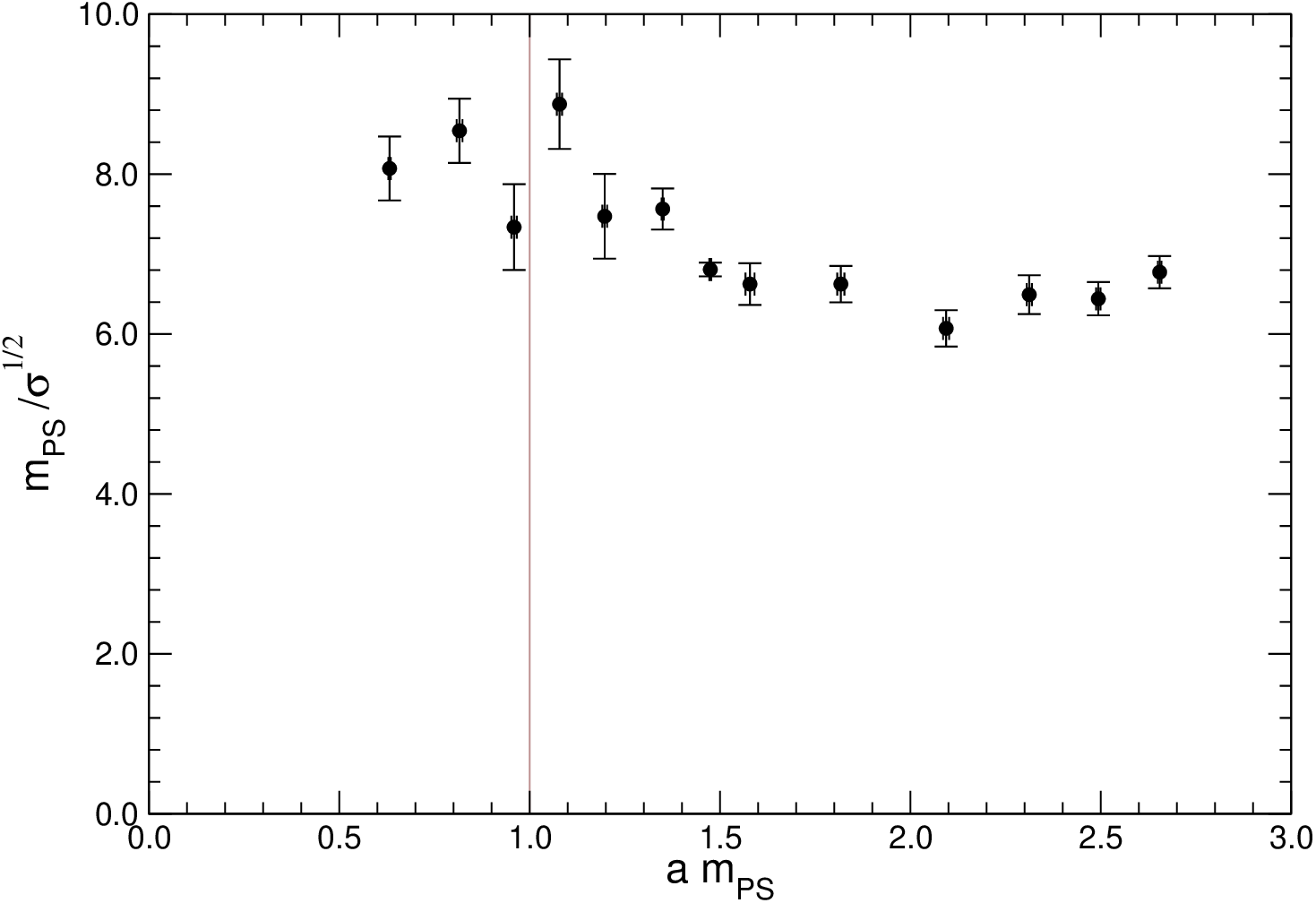}
\caption{Pseudoscalar mass in units of $\sqrt{\sigma}$ as a function
  of the pseudoscalar mass in units of $a^{-1}$. Points on the left of
  the vertical line at $a M = 1$ are expected to be reasonably free
  from finite lattice spacing effects.}
\label{fig:mpsoversigma}
\end{figure}

{\bf A - Hierarchy in the spectrum.} A global overview of our
numerical results is presented in Fig.~\ref{fig:all}. The plot shows
that a small, but clearly non--vanishing string tension exists at
least down to PCAC masses of the order of 0.1 in units of the inverse
lattice spacing, and that such string tension decreases with
decreasing PCAC mass. The string tension shown in the figure is
extracted using correlators of Polyakov loops, but compatible numbers
are obtained from the static potential. For a lattice of fixed size
$L$, the string tension decreases when decreasing the PCAC mass until it
becomes ${\cal O}(1/L^2)$ at which value a plateaux is reached. Tests on lattices of
different sizes show that this is a finite size effect. As the lattice
size is increased, the string tension keeps decreasing as the fermion
mass is decreased. A non--zero string tension is expected in the
massive case even in the conformal window, since a non--zero quark mass
moves the theory away from the attraction basin of the IR fixed
point. What is remarkable is that even at our lowest PCAC masses there
is a well--defined hierarchy in the spectrum: the string tension
defines the lowest mass scale in the system, and the meson spectrum is
well above the lowest--lying glueballs. At this stage, it is worth
noticing that states with mass of the order or above $a^{-1}$ are
expected to be significantly affected by discretization artifacts;
while we can reach small masses for the mesons, the extraction of the
gluonic spectrum becomes very expensive for light fermions. As a
consequence we do not have results for the glueballs and the string
tension at the smaller values of the mass. Despite the fact that a significant
portion of our spectrum falls in the region were discretization
artifacts are not under control, the hierarchy of the spectrum seems
to be a robust conclusion, as it can be extrapolated smoothly to the
region where discretization errors are expected to be under
control. To investigate in more detail the observed hierarchy of
scales, we plot in Fig.~\ref{fig:mpsoversigma} the ratio
$m_\mathrm{PS}/\sqrt{\sigma}$ as a function of the pseudoscalar mass
in lattice units. For a standard confining and chiral symmetry
breaking theory, this ratio goes to zero in the chiral limit. For our
theory, even when varying the pseudoscalar mass by a factor of six to
a region where it is well below the cutoff scale, this ratio is always
of order 10, and does not extrapolate to zero in the chiral
limit. This behavior is at odds with the one expected for QCD, and
indeed it is not observed in QCD simulations for similar variations of
the pseudoscalar mass. Fermion loops seem to strongly affect the
gluonic sector, keeping the corresponding scale always well below the
scale of mesonic physics.
 Let us discuss now the main features emerging from the numerical
calculations.\\ 
\begin{figure}[bh]
\includegraphics*[scale=0.35]{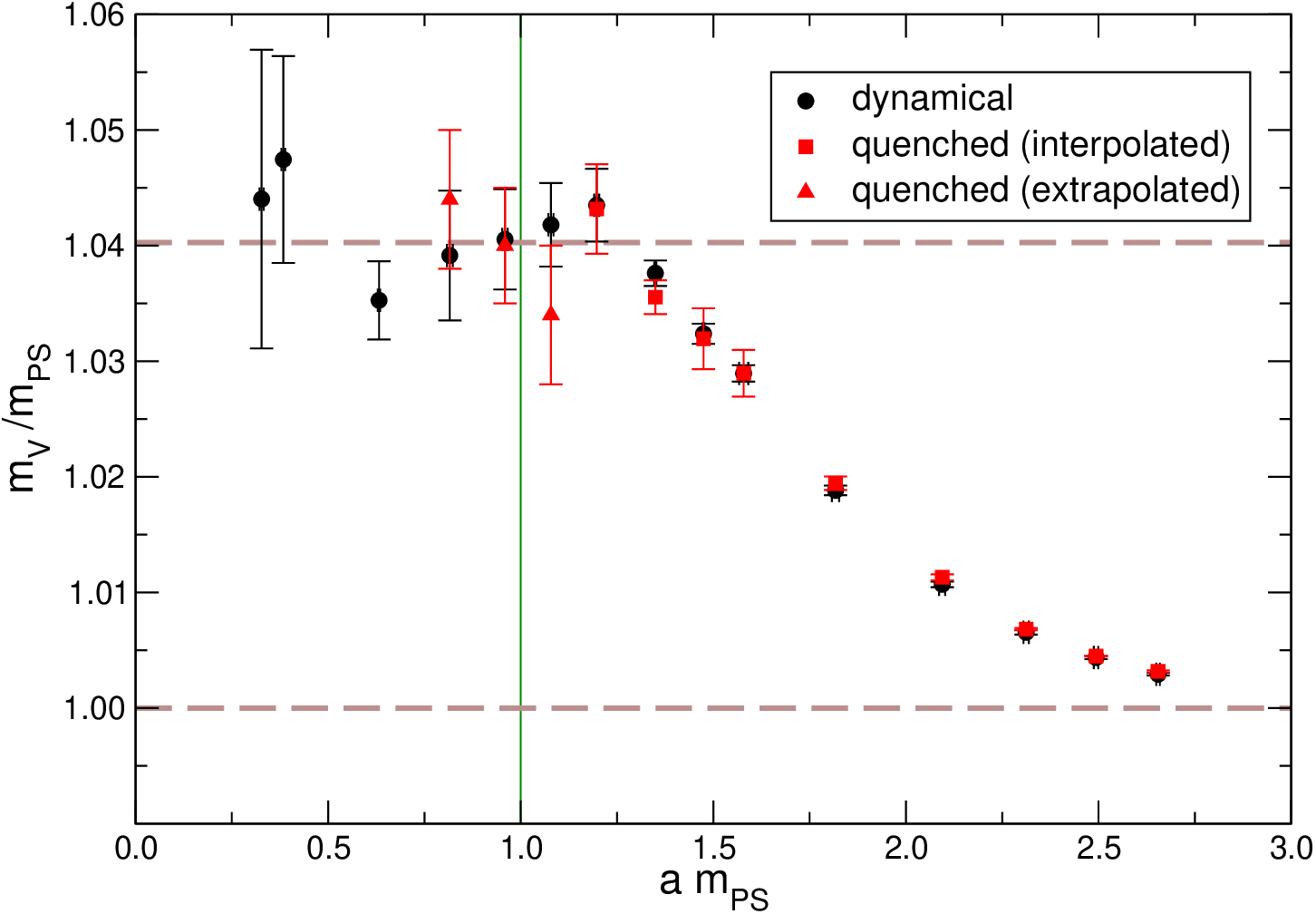}
\caption{The ratio of the vector mass $m_V$ over the pseudoscalar mass
  $m_\mathrm{PS}$ as a function of the pseudoscalar mass. Quenched
  data at equivalent bare lattice parameters are also shown.}
\label{fig:quenched3}
\end{figure}
\begin{figure}[ht]
\includegraphics*[scale=0.35]{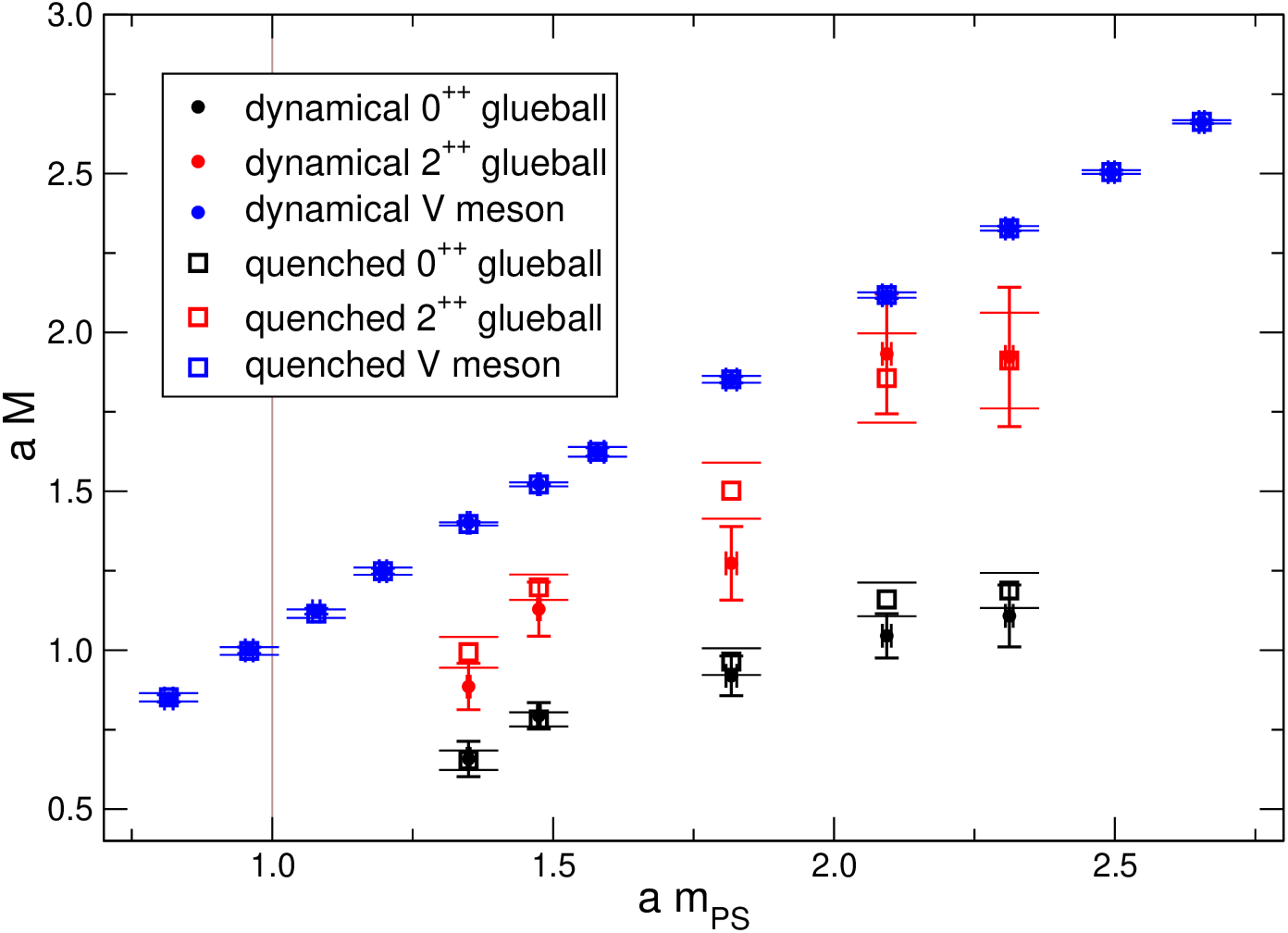}
\caption{Comparison of the lightest pseudoscalar and glueball masses in the quenched and in the
  dynamical theory as a function of the pseudoscalar mass.}
\label{fig:quenched2}
\end{figure}
~\\ 
{\bf B - IR effective dynamics.} The vector meson is not displayed in
Fig.~\ref{fig:all}, since it is degenerate with the pseudoscalar meson
on the scale of the plot (the approximate degeneracy of the
pseudoscalar and vector mesons was already observed in previous
simulations, starting from Ref.~\cite{Catterall:2007yx}). Having a
better control on the massless limit allows us to investigate more
closely the degeneracy between the pseudoscalar and the vector meson
observed in previous studies. Our data in Fig.~\ref{fig:quenched3}
show that, as the fermion mass is reduced, the ratio
$m_\mathrm{V}/m_\mathrm{PS}$ progressively rises from 1 (which is
the expected result in the heavy quark effective theory) to 1.04,
where it seems to stabilize. Understanding whether this 4\% variation
is significant will require a more systematic study. To shed more
light on this ratio, we performed a comparison with data obtained in
quenched SU(2) simulations. The bare coupling and the fermion mass of
the quenched theory need to be fine--tuned in such a way that the
string tension and the pseudoscalar mass of the dynamical simulations
are matched. 
The comparison in Fig.~\ref{fig:quenched3}
shows that the splitting between the pseudoscalar and vector meson
masses is due purely to gluonic effects. To see how much the spectrum
resembles the pure Yang--Mills one, we have compared in
Fig.~\ref{fig:quenched2} the glueball masses obtained from our
simulations to the ones of pure gauge SU(2) after matching the string
tensions. Data for the Yang-Mills theory are obtained by interpolating
(or slightly extrapolating) the data in Ref.~\cite{Lucini:2004my}.
Once the string tension and the pseudoscalar mass are tuned,
all the states that we studied are reproduced by the quenched
data. Note that the value of the bare lattice coupling $\beta$ needs
to be adjusted as a function of the fermion mass used in the dynamical
simulations. A possible explanation of this behavior is that the
mesons decouple in the infrared, and the effective long--distance
theory is SU(2) Yang-Mills with a hadronic scale smaller than the
fermion mass of the full theory.

Another explanation for the features observed in the mesonic spectrum
could be that the fermions are simply too massive and as a consequence
the theory is quenched; however, this conclusion is unlikely to
accommodate the observed hierarchy between mesonic and gluonic states
in the spectrum, which seems to reflect the importance of fermion
loops in the UV. In fact, the simultaneous presence of these two
phenomena could be seen as a signature of a conformal point in the
massless limit. Such a scenario has been proposed by
Miransky~\cite{Miransky:1998dh}, who has shown that in SU($N$) gauge
theories with fundamental fermions in the conformal phase but close to
the higher end of the conformal window (where the infrared fixed point
is perturbative) the low--energy effective spectrum coincides with the
spectrum of the SU($N$) Yang--Mills theory with a dynamically generated
mass scale that is proportional to the pseudoscalar meson mass. The
proportionality constant is exponentially small in the inverse of the
squared gauge coupling at the IR fixed point. While the details of the
calculation can not be trusted at the lower end of the conformal
window (in which our theory would be if an infrared fixed point
existed), it is conceivable that features like the hierarchy in the
spectrum and the suppressed infrared scale will survive also when a
perturbative analysis of the physics near the infrared fixed point is
not reliable.

In conclusion, our numerical data are the first systematic study of
the chiral regime for the SU(2) gauge theory with two Dirac adjoint
fermions. They support a scenario in which the spectrum is determined
by a pure gauge dynamics, whose dynamically--generated scale {\em
slides} with the fermion mass and is always well below the scale of
the meson physics (the separation being about one order of
magnitude). Moreover all the states in the mesonic spectrum become
lighter as the fermion mass is decreased; the separation
of the pseudoscalar Goldstone bosons from the rest of the spectrum, which would characterize the
spontaneous breaking of chiral symmetry, is not observed.  These peculiar features of
the spectrum provide a stronger evidence in favor of the conformality
of the theory in the massless limit.

Indeed if the behavior observed over the limited range of quark masses in the
scaling region of our present simulations does extrapolate to the continuum and
chiral limit without any significant qualitative difference, the conclusion
that this theory lies in the conformal window will be natural.

At the moment, however our conclusions are limited by a number of factors:
gluonic observables are very expensive to measure accurately at small 
quark masses so at present we only have two points with glueball masses below
the cutoff scale.
Moreover finite volume effects have to be under control even for the mesonic
spectrum to be extrapolated with confidence to the small mass regime.  This
makes it very difficult to get closer to the chiral limit, and explicitly check
that we are not in a heavy quark regime.
Finally, we  lack control on the continuum limit, since all of our numerical simulations
were performed at a single lattice spacing.
Scaling towards the continuum limit is beyond the scope of the present paper and it will be addressed by forthcoming simulations.

In view of these limitations a QCD-like scenario still remains possible, as well as 
the intermediate Walking scenario. At present however, taking into account the 
available information presented in this paper as well as in previous publications, 
the alternative which is more likely is that this theory is IR conformal.
To put this statement on solid grounds, more accurate lattice studies are 
still necessary.   

The discussion presented in this paper outlines a strategy to
understand the phase of the massless theory in the continuum limit.
Simulations will be extended to other values of the lattice spacing
$a$, and smaller pseudoscalar masses, with the twofold aim to
extrapolate the results to the continuum and to keep under control
cutoff effects on the spectrum. This should allow us to resolve the
issues related to the possibility of our constituent fermions being
still too heavy, and of our spectrum being influenced by an IR fixed
point not related to the continuum theory. The comparison of the
gluonic and the mesonic sector looks like a promising way to address
the issue of conformality by lattice simulations. Finally, following
the recent numerical studies of the conformal window in SU(3) gauge
theory with fermions in the fundamental
representation~\cite{Appelquist:2007hu,Deuzeman:2008sc,Deuzeman:2009mh,Appelquist:2009ty,Hasenfratz:2009ea},
we notice that applying a combined analysis of the meson and glueball
spectrum to that theory at both ends of the conformal phase could help
to clarify the physics of theories with IR fixed points.

~\\
We are indebted with C.~Allton, T.~DeGrand, A.~Hasenfratz, M.~Piai, and
F.~Sannino for insightful discussions. This work has been partially supported
by STFC under contracts PP/E007228/1 and ST/G000506/1. BL is supported by the
Royal Society. LDD is supported by an STFC advanced fellowship.  Definite
progress towards the completion of this work was made during the fruitful
Large--N conference in Swansea.  Most of the numerical results presented in
this work have been obtained on the BlueC supercomputer at Swansea University.

\bibliographystyle{apsrev}
\bibliography{twoscales}
\end{document}